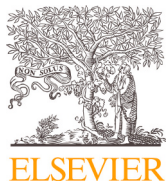

Contents lists available at ScienceDirect

## Nuclear Physics, Section B

journal homepage: www.elsevier.com/locate/nuclphysb

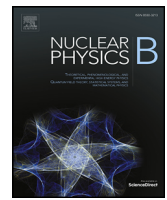

High Energy Physics - Theory

# Breakthrough on dynamical higgs mechanism for dRGT gravity: Example in which graviton gains mass through electroweak phase transition

Emmanuel Kanambaye [1]

*Librairie BAH SARL, Hall du Grand Hôtel de Bamako-Mali, BP 104, Bamako, Mali*

ABSTRACT

Massive gravity is an interesting theoretical framework capable of explaining among others things cosmological IR (infra-red) phenomena like late-time cosmic acceleration.

Unfortunately, it turned out to be very difficult of getting consistent massive gravity theory due, among other things, to ghost appearance and strong coupling break-down problems.

Of course, even though since the work of de Rham, Gabadadge and Tolley; the ghost problem appears overcame; things remain unchanged for the strong coupling break-down problem which is the problem that a massive gravity of experimentally viable graviton mass $m$ has (in comparison of standard massless gravity theory) a very low cutoff-scale $\Lambda_3 = [m^2 M_P]^{\frac{1}{3}}$ above which the theory fails; a problem what must be resolved if we want massive gravity be a consistent physical theory which can pretend to describe nature.

Now as we know, one of the better way of overcoming this strong coupling break-down problem of massive gravity is to have a dynamical Higgs mechanism for gravity capable of providing a clear-cut way of making the graviton massless dynamically above the cutoff-scale $\Lambda_3$; a challenge what turned out to be difficult to surmount.

It is this difficult that I overcome in the present paper by showing that it is quite possible of getting a four-dimensional modified massless gravity theory which becomes dynamically massive through for example the dynamical standard model electroweak phase transition.

More precisely, I propose the first example of dynamical Higgs mechanism for dRGT gravity permitting to overcome the strong coupling break-down problem of dRGT gravity.

## 1. Introduction

After Fierz and Pauli work [1], the question of whether gravity can be made dynamically massive in a consistent way has roughly always arisen.
Indeed contrary to interactions mediated by vector fields which could be made dynamically massive in a consistent way through a spontaneous symmetry breaking of the kind Higgs mechanism [2–4]; gravitation gives generically rise to difficulty.
In fact, even to have a ghost free Lorentz Invariant massive gravity has been a challenge until de Rham, Gabadadze and Tolley find acceptable trick [5].

*E-mail address:* wadouba@gmail.com

[1] Independent Researcher

Nevertheless, in spite of the advance achieved by de Rham, Gabadadze and Tolley [5], another major problem is weighing down massive gravity, the so-called strong coupling break-down problem of massive gravity [6,7].

Indeed the strong coupling break-down problem of dRGT massive gravity is the problem that, in comparison to standard massless gravity theory [8], dRGT gravity breaks-down above a very low cutoff-scale $\Lambda_3 = [m^2 M_P]^{\frac{1}{3}}$ for experimentally viable graviton mass $m$ [6,7]; whence it is necessary to overcome this problem if we want that massive gravity be a consistent physical theory which can pretend to describe nature.

Of course as we know [6,7,9]; one of the better way of overcoming the strong coupling break down problem of massive gravity is to find a consistent dynamical higgs mechanism for gravity through which the graviton acquires a mass $m > 0$ below the characteristic scale $\Lambda_3 = [m^2 M_P]^{\frac{1}{3}}$.

Indeed in such a case, the strong coupling break-down problem of massive gravity would be handled since the theory would become massless in a consistent way above the strong coupling break-down scale $\Lambda_3$ and thus would have a massless gravity cutoff-scale (generally expected to be of Planck mass $M_p \sim 10^{19} GeV$ scale) unless the gotten massless gravity be UV complete.

Unfortunately, as we know [6,7,9]; it turned out to be roughly difficult to find "reasonable" trick permitting such an advance, and it is why, until now [6,9,10], massive gravity was "always" formulated as an effective field theory with a cutoff-scale $\Lambda_3$ much smaller than of standard massless gravity [8].

In the present paper, it is this difficulty that I propose to overcome; more precisely by involving some auxiliary fields and Lagrange multipliers; I succeeded to achieve from variational principle, a consistent massless gravity theory which becomes (dynamically) massive through for example the standard model [11] electroweak phase transition [2–4].

More exactly, I achieve the first example of dynamical Higgs mechanism for dRGT gravity permitting to circumvent the strong coupling break-down problem of massive gravity.

Before to start, let's emphasize that our aim is to show that it is in principle, quite possible to get/construct from variational principle a consistent massless gravitational theory which (irrespective of its justification from the point of view of some underlying more fundamental physics or principle) become dynamically massive through a dynamical phase transition; as consequence we will in the present paper, not try to justify or motivate the action/lagrangian which yields our theory from the point of view of any underlying more fundamental physics or principle like quantum gravity or other which anyway is not known yet.

The paper is organised as follow, in the Section 2, I make a brief recall of original dRGT massive gravity [5] and Higgs mechanism basic [2–4]; In the section 3, I derive the model that I propose from variational principle; In the section 4, I show that the model is a massive gravity after electroweak phase transition and massless before electroweak phase transition; in the Section 5 I show how the model could escape to the strong coupling break-down problem of massive gravity, while I devote Section 6 for conclusion.

The signature convention ( - + + + ) will be assumed for the metric; likewise Einstein's summation convention will be assumed.

## 2. Brief recall

### 2.1. Brief recall of dRGT massive gravity

In [5], de Rham, Gabadadze and Tolley proposed among the simplest ghost-free Lorentz Invariant massive gravity from the action:

$$S = \frac{M_P^2}{2} \int \left[ R + 2\kappa L_M + m^2 \mathcal{U}(g, \varphi^\rho) \right] \sqrt{-g} d^4 x \tag{1}$$

where $M_P$ denotes Planck mass; $R$ the Ricci scalar; $\kappa$ the Einstein gravitational constant; $L_M$ the lagrangian of standard matter [11]; $m$ a constant; $\varphi^\rho = (\varphi^0, \varphi^j)$ a four-components Stückelberg scalar field, while [5,10]:

$$\mathcal{U}(g, \varphi^\rho) = \mathcal{U}_2 + \alpha_3 \mathcal{U}_3 + \alpha_4 \mathcal{U}_4 \tag{2}$$

$$\mathcal{U}_2 = [K]^2 - [K^2] \tag{3}$$

$$\mathcal{U}_3 = [K]^3 - 3[K][K^2] + 2[K^3] \tag{4}$$

$$\mathcal{U}_4 = [K]^4 - 6[K^2][K^2] + 8[K][K^3] + 3[K^2]^2 - 6[K^4] \tag{5}$$

$$K_\nu^\mu = \delta_\nu^\mu - M_\nu^\mu \tag{6}$$

$$M_\sigma^\mu M_\nu^\sigma = g^{\mu\sigma} f_{\sigma\nu} = g^{\mu\sigma} \partial_\sigma \varphi^\rho \partial_\nu \varphi^\beta f_{\rho\beta} \tag{7}$$

with $[K] = K_\mu^\mu$; $[K^2] = K_\nu^\mu K_\mu^\nu$; $[K^3] = K_\nu^\mu K_\rho^\nu K_\mu^\rho$; $[K^4] = K_\nu^\mu K_\rho^\nu K_\sigma^\rho K_\mu^\sigma$; $f_{\rho\beta}$ a fudicial metric that can be Minkowski metric, while $\alpha_3$ and $\alpha_4$ are free constants that can be zero.

Indeed the action (1) would yield the famous dRGT massive gravity field equations [10]:

$$G_{\mu\nu} = R_{\mu\nu} - \frac{1}{2} R g_{\mu\nu} = \kappa T_{\mu\nu} - m^2 \mathcal{U}_{\mu\nu} \tag{8}$$

$$\mathcal{E}_\nu = \partial_\mu \mathcal{U}_\nu^\mu = \nabla_\mu \left[ \frac{\partial \mathcal{U}(g, \varphi^\rho)}{\partial(\partial_\mu \varphi^\rho)} \right] = \nabla^\mu \mathcal{U}_{\mu\nu} = 0 \tag{9}$$

where $T_{\mu\nu} = -\frac{2}{\sqrt{-g}}\frac{\delta(\sqrt{-g}L_M)}{\delta g^{\mu\nu}}$ denotes the energy-momentum tensor of standard matter, and (9) the equation of motion for the Stückelberg fields $\varphi^\rho$, while [5,9,10]:

$$\mathcal{U}_{\mu\nu} = \frac{\delta(\mathcal{U}(g,\varphi^\rho)\sqrt{-g})}{\sqrt{-g}\delta g^{\mu\nu}} = K_{\mu\nu} - [K]g_{\mu\nu} - [1+3\alpha_3]\left[K_{\mu\nu}^2 - [K]K_{\mu\nu} - \frac{\mathcal{U}_2}{2}g_{\mu\nu}\right] + 3[4\alpha_4 + \alpha_3]\left[K_{\mu\nu}^3 - [K]K_{\mu\nu}^2 + \frac{\mathcal{U}_2}{2}K_{\mu\nu} - \frac{\mathcal{U}_3}{6}g_{\mu\nu}\right] \qquad (10)$$

Of course, we are not going to do a review of dRGT model here (we recomand the papers [5,9,10] to reader interested by further details on the model); nevertheless it is worth to emphasize that dRGT gravity is among the first example of ghost free Lorentz-Invariant massive gravity; unfortunately, it fails above the strong-coupling scale $\Lambda_3 = [M_P m^2]^{\frac{1}{3}}$ well below Planck scale for any experimentally viable graviton mass $m > 0$.

This mainly comes from the fact that the model is not achieved as a massless gravity theory which becomes dynamically massive through a dynamical phase transition endowing the graviton with mass below the characteristic energy-scale $\Lambda_3 = [M_P m^2]^{\frac{1}{3}}$; a goal proved to be difficult to achieve [6,7,9].

In the next sections, we are going to see how this difficulty can be overcame.

But before, let's make a brief recall of the basic of standard model Higgs mechanism [2–4].

### 2.2. Brief recall on electroweak phase transition

As we well know; the Higgs field $\Phi$ of the standard model is a doublet of complex scalar fields $\phi^+ = \frac{\varphi_1 + i\varphi_2}{\sqrt{2}}$ and $\phi^0 = \frac{\varphi_3 + i\varphi_4}{\sqrt{2}}$, charged under the $SU(2)_L \times U(1)_Y$ group characterized by two coupling constants $g$ and $g'$ [12] i.e.:

$$\Phi = \begin{pmatrix} \phi^+ \\ \phi^0 \end{pmatrix} = \frac{1}{\sqrt{2}}\begin{pmatrix} \varphi_1 + i\varphi_2 \\ \varphi_3 + i\varphi_4 \end{pmatrix} \qquad (11)$$

The lagrangian of this Higgs field $\Phi$ reads [2–4,12]:

$$L_H = -[D_\mu \Phi]^\dagger [D^\mu \Phi] - V(\Phi^\dagger, \Phi) \qquad (12)$$

where:

$$D_\mu = \partial_\mu + ig\frac{Y_H}{2}B_\mu + ig'\tau_a W_\mu^a \qquad (13)$$

with $B_\mu$ and $W_\mu^a$ the electroweak gauge bosons; $Y_H$ the Higgs weak hypercharge and $\tau_a$ the generators of the $SU(2)$ group; also the index $\mu$ runs from 0 to 3 i.e. $\mu = 0, 1, 2, 3$, while the index $a$ runs from 1 to 3 i.e. $a = 1, 2, 3$.

Likewise usually, the potential $V(\Phi^\dagger, \Phi)$ is chosen to be:

$$V(\Phi^\dagger, \Phi) = m_0^2 \Phi^\dagger \Phi + \lambda[\Phi^\dagger \Phi]^2 \qquad (14)$$

where $m_0$ denotes the effective mass parameter of $\Phi$ and $\lambda > 0$ a positive constant.

Indeed, the interest of this above potential (14) is that for $m_0^2 < 0$, there will be a spontaneous symmetry breaking endowing the field $\Phi$ with a non-zero vacuum expectation value [2–4,12]:

$$<\Phi> = <\phi^0> = \frac{1}{\sqrt{2}}v = \sqrt{-\frac{m_0^2}{2\lambda}} \qquad (15)$$

As consequence, the perturbation of $\Phi$ around its minimum could be written as [2–4,12]:

$$\Phi = \frac{1}{\sqrt{2}}(v+H)\begin{pmatrix} 0 \\ 1 \end{pmatrix} = \frac{1}{\sqrt{2}}\begin{pmatrix} 0 \\ v+H \end{pmatrix} \qquad (16)$$

where $H$ is a real scalar field, the Higgs boson! what would give [2–4,12]:

$$\Phi^\dagger \Phi = \frac{1}{2}[v+H]^2 = \frac{1}{2}v^2 + \frac{1}{2}[2v+H]H \qquad (17)$$

whence the lagrangian (12) would become [2–4,12]:

$$L_H = -\frac{1}{2}\partial_\mu H \partial^\mu H - V(H) - \frac{1}{4}[v^2 + 2vH + H^2]\left[g'^2 W_\mu^+ W^{-\mu} + \frac{m_Z^2}{v^2}Z_\mu Z^\mu\right] \qquad (18)$$

where [2–4,12]:

$$V(H) = \lambda v^2 H^2 + \lambda v H^3 + \frac{\lambda}{4}H^4 - \frac{\lambda v^4}{4} \qquad (19)$$

while $Z_\mu$ and $W_\mu^\pm$ are respectively the electrically neutral and charged weak bosons of masses $m_Z = \frac{v}{\sqrt{2}}\sqrt{g^2 + g'^2}$ and $m_W = \frac{v}{\sqrt{2}}g'$; whence the assertion that Higgs field $\Phi$ yields, through dynamical spontaneous electroweak symmetry breaking, masses for the weak interaction bosons $Z_\mu$ and $W_\mu^{\pm}$".

Of course to capture the whole Higgs mechanism, we have to consider amongst other things, Yukawa coupling of the Higgs field with fermions; Kinetic term of the gauge bosons etc…[2–4,12].

*2.3. Terms necessary to our trick*

Before starting the next section; let's introduce the special function $k(\Phi)$ of the standard model Higgs doublet $\Phi$, reading:

$$k(\Phi) = \left[2B(e^{\Phi_1}, 1 + e^{\Phi_2}) - B(e^{\Phi_2}, e^{\Phi_2})\right]^2 \qquad (20)$$

where e *is the Euler's number or Napier's constant;* $B(x,y) = \frac{\Gamma(x)\Gamma(y)}{\Gamma(x+y)}$ *the beta function;* $\Phi_1 = \frac{1}{4v}\sqrt{v^2 - [2v - \sqrt{2\Phi^\dagger \Phi}]\sqrt{\Phi^{\dagger *}\Phi + \Phi^\dagger \Phi^*}}$ *and* $\Phi_2 = \frac{\sqrt{2\Phi^\dagger \Phi}}{4v} - \frac{1}{4}$, *with* $v$ *the constant given by* (15); $\Phi^*$ *the complex conjugate of* $\Phi$ *and* $\Phi^{\dagger *}$ *the complex conjugate of* $\Phi^\dagger$.

Of course, we can verify that by constuction of $\Phi_1$ and $\Phi_2$; it is only when the full SU(2) symmetry associated to $\Phi$ breaks, like in the case of (16) where $\Phi$ develops the non-zero vacuum expectation value $\frac{v}{\sqrt{2}}$, that we could have:

$$\Phi_1 = \Phi_2 = \frac{H}{4v} \qquad (21)$$

Now since the beta function satisfies the identities [13]:

$$B(x, 1+y) = \frac{y}{x+y}B(x,y) \qquad (22)$$

$$2B(y, 1+y) = B(y,y) \qquad (23)$$

then it results that we will have $k(\Phi) = 0$ only in the case of (21) where $\Phi_1 = \Phi_2 = \frac{H}{4v}$; more precisely only when $\Phi \neq 0$ with at least $\varphi_2 = \varphi_4 = 0$ like in the case of (16). Said otherwise, by construction, the special function $k(\Phi)$ given by (20), becomes zero only when the full SU(2) symmetry breaks i.e. only after the electroweak phase transition; what makes $k(\Phi)$ a really interesting special function. Now, by writting the fudicial metric $f_{\rho\beta}$ as:

$$f_{\rho\beta} = \begin{pmatrix} f_{00} & f_{01} & f_{02} & f_{03} \\ f_{10} & f_{11} & f_{12} & f_{13} \\ f_{20} & f_{21} & f_{22} & f_{23} \\ f_{30} & f_{31} & f_{32} & f_{33} \end{pmatrix} \qquad (24)$$

we can, by using the Stückelberg scalar fields $\varphi_\rho$ and the function $k(\Phi)$ defined above, construct a new tensor $\Phi_{\rho\beta}$ defined for example as:

$$\Phi_{\rho\beta} = \begin{pmatrix} e^{-[\varphi_0 k(\Phi)\varphi_0]}f_{00} & e^{-[\varphi_0 k(\Phi)\varphi_1]}f_{01} & e^{-[\varphi_0 k(\Phi)\varphi_2]}f_{02} & e^{-[\varphi_0 k(\Phi)\varphi_3]}f_{03} \\ e^{-[\varphi_1 k(\Phi)\varphi_0]}f_{10} & e^{-[\varphi_1 k(\Phi)\varphi_1]}f_{11} & e^{-[\varphi_1 k(\Phi)\varphi_2]}f_{12} & e^{-[\varphi_1 k(\Phi)\varphi_3]}f_{13} \\ e^{-[\varphi_2 k(\Phi)\varphi_0]}f_{20} & e^{-[\varphi_2 k(\Phi)\varphi_1]}f_{21} & e^{-[\varphi_2 k(\Phi)\varphi_2]}f_{22} & e^{-[\varphi_2 k(\Phi)\varphi_3]}f_{23} \\ e^{-[\varphi_3 k(\Phi)\varphi_0]}f_{30} & e^{-[\varphi_3 k(\Phi)\varphi_1]}f_{31} & e^{-[\varphi_3 k(\Phi)\varphi_2]}f_{32} & e^{-[\varphi_3 k(\Phi)\varphi_3]}f_{33} \end{pmatrix} \qquad (25)$$

where e *is the Euler's number, and* $f_{00}$, $f_{01}$ *etc…the components of the fudicial metric* $f_{\rho\beta}$. Now what is interesting with this tensor $\Phi_{\rho\beta}$, is that it becomes the fudicial metric tensor $f_{\rho\beta}$ when $k(\Phi) = 0$, because in the case of $k(\Phi) = 0$, we will have:

$$\Phi_{\rho\beta} = \begin{pmatrix} e^0 f_{00} & e^0 f_{01} & e^0 f_{02} & e^0 f_{03} \\ e^0 f_{10} & e^0 f_{11} & e^0 f_{12} & e^0 f_{13} \\ e^0 f_{20} & e^0 f_{21} & e^0 f_{22} & e^0 f_{23} \\ e^0 f_{30} & e^0 f_{31} & e^0 f_{32} & e^0 f_{33} \end{pmatrix} = \begin{pmatrix} f_{00} & f_{01} & f_{02} & f_{03} \\ f_{10} & f_{11} & f_{12} & f_{13} \\ f_{20} & f_{21} & f_{22} & f_{23} \\ f_{30} & f_{31} & f_{32} & f_{33} \end{pmatrix} \qquad (26)$$

while, in the case of $k(\Phi) \neq 0$; $\Phi_{\rho\beta}$ will not be a fudicial metric tensor, since incapable of providing a fixed reference background metric alongside the physical spacetime metric $g_{\mu\nu}$. Said otherwise, if in the case of $k(\Phi) = 0$, the tensor $\Phi_{\rho\beta}$ becomes the fudicial metric tensor $f_{\rho\beta}$; in the case of $k(\Phi) \neq 0$, $\Phi_{\rho\beta}$ is not a fudicial metric tensor at all.

Thus for example, by replacing into the tensor $\mathcal{U}_{\mu\nu}$ given by (10), the fudicial metric $f_{\rho\beta}$ by the tensor $\Phi_{\rho\beta}$ given by (25), we will get the tensor $\mathcal{U}_{\mu\nu}(\Phi)$ reading:

$$\mathcal{U}_{\mu\nu}(\Phi) = K_{\mu\nu} - [K]g_{\mu\nu} - [1+3\alpha_3]\left[K_{\mu\nu}^2 - [K]K_{\mu\nu} - \frac{\mathcal{U}_2}{2}g_{\mu\nu}\right] + 3[4\alpha_4 + \alpha_3]\left[K_{\mu\nu}^3 - [K]K_{\mu\nu}^2 + \frac{\mathcal{U}_2}{2}K_{\mu\nu} - \frac{\mathcal{U}_3}{6}g_{\mu\nu}\right] \qquad (27)$$

where $K_\nu^\mu = \delta_\nu^\mu - M_\nu^\mu$, *and* $M_\sigma^\mu M_\nu^\sigma = g^{\mu\sigma}\Phi_{\sigma\nu} = g^{\mu\sigma}\partial_\sigma\varphi^\rho\partial_\nu\varphi^\beta \Phi_{\rho\beta}$, *while* $[K] = K_\mu^\mu$; $K_{\mu\nu}^2 = K_{\mu\sigma}K_\nu^\sigma$; $\mathcal{U}_2 = [K]^2 - [K^2]$; $\mathcal{U}_3 = [K]^3 - 3[K][K^2] + 2[K^3]$; $[K^2] = K_\nu^\mu K_\mu^\nu$ *etc…*

This means that the tensor $\mathcal{V}'_{\mu\nu}(\Phi)$ will become the tensor $\mathcal{V}_{\mu\nu}$ when $k(\Phi) = 0$, while when $k(\Phi) \neq 0$; $\mathcal{V}'_{\mu\nu}(\Phi)$ will be a tensor not involving a fudicial metric, more precisely a tensor incapable of providing a fixed reference background metric alongside the physical spacetime metric $g_{\mu\nu}$. As far as we are concerned, $k(\Phi)$ and $\mathcal{V}'_{\mu\nu}(\Phi)$ are enough/necessary to understand the rest of the paper.

## 3. A new gravity theory: Derivation from variational principle

To get the paradigm allowing gravity to become dynamically massive through electroweak phase transition; let's begin by modifying the action (1) as follow:

$$S = \frac{M_P^2}{2} \int \left[ R + 2\kappa L_M + m^2 \mathcal{U}(g, \varphi^\rho) - L_{\chi_0} - L_{\chi_1} - L_{\chi_2} - \frac{\chi_2}{m} [\mathcal{E}_{\mu\nu} \mathcal{U}^{\mu\nu}]^2 \right] \sqrt{-g} d^4 x \tag{28}$$

1) where $L_{\chi_0}$, $L_{\chi_1}$ and $L_{\chi_2}$ are additional lagrangians respectively reading:

$$L_{\chi_0} = \omega_0 \left[ \partial_\mu \chi_0 \partial^\mu \chi_0 + 1 \right] \tag{29}$$

$$L_{\chi_1} = \omega_1 \left[ \partial_\mu \chi_1 \partial^\mu \chi_1 + 1 \right] + \varepsilon_1 \left[ \omega_1 - m^2 \right] - m^2 \partial_\mu \chi_1 \partial^\mu \chi_1 \tag{30}$$

$$L_{\chi_2} = \omega_2 \left[ \partial_\mu \chi_2 \partial^\mu \chi_2 + 1 \right] + \varepsilon_2 \left[ \omega_2 + m^2 \right] + m^2 \partial_\mu \chi_2 \partial^\mu \chi_2 \tag{31}$$

in which $\omega_0$, $\chi_0$, $\omega_1$, $\chi_1$, $\varepsilon_1$, $\omega_2$, $\chi_2$ and $\varepsilon_2$ are auxiliary variable, with $m$ a dimensioned free constant.

2) while $\mathcal{U}_{\mu\nu}$ is the tensor given by (10) and $\mathcal{E}_{\mu\nu}$ a tensor reading:

$$\mathcal{E}_{\mu\nu} = \omega_0 \partial_\mu \chi_0 \partial_\nu \chi_0 - m^2 \mathcal{U}_{\mu\nu} - k(\Phi) A_\mu A_\nu - k(\Phi) S_{\mu\nu}(g, \Psi, \varphi^\rho, A) + m^2 \mathcal{V}'_{\mu\nu}(\Phi) \tag{32}$$

where $A_\mu$ is an auxiliary vector field; $k(\Phi)$ the function given by (20); $\mathcal{V}'_{\mu\nu}(\Phi)$ the tensor given by (27) and $S_{\mu\nu}(g, \Psi, \varphi^\rho, A)$ a free symmetric tensor that can depend on the standard matter fields, the metric, the Stückelberg fields $\varphi^\rho$ and/or their derivatives etc..., as well as of the auxiliary vector field $A_\mu$ i.e. $S_{\mu\nu}(g, \Psi, \varphi^\rho, A)$ must not involves derivatives of $A_\mu$; of course we can quite assume case $S_{\mu\nu}(g, \Psi, \varphi^\rho, A) = 0$.

It is worth mentioning that we consider the terms $k(\Phi)$ and $\mathcal{V}'_{\mu\nu}(\Phi)$ only because of their ability to yield the tensor $\mathcal{U}_{\mu\nu}$ after electroweak phase transition; we will give no other justification of their use insofar as our aim is not to justify (for the moment) our action (28) from the point of view of some underlying more fundamental physics or principle; we just require that it be a consistent/stable action; which is the case.

Now as all the elements of the additional lagrangians (29)–(31) and tensor $\mathcal{E}_{\mu\nu}$ are specified; then we can apply variational principle; indeed:
a) varying the action (28) with respect to $\delta g^{\mu\nu}$ would yield:

$$G_{\mu\nu} = \kappa T_{\mu\nu} - m^2 \mathcal{U}_{\mu\nu} + \frac{1}{\sqrt{-g}} \left[ \frac{\delta(\sqrt{-g} L_{\chi_0})}{\delta g^{\mu\nu}} + \frac{\delta(\sqrt{-g} L_{\chi_1})}{\delta g^{\mu\nu}} + \frac{\delta(\sqrt{-g} L_{\chi_2})}{\delta g^{\mu\nu}} + \frac{\delta(\sqrt{-g} \frac{\chi_2}{m} [\mathcal{E}_{\mu\nu} \mathcal{U}^{\mu\nu}]^2)}{\delta g^{\mu\nu}} \right] \tag{33}$$

where of course $T_{\mu\nu} = -\frac{2}{\sqrt{-g}} \frac{\delta(\sqrt{-g} L_M)}{\delta g^{\mu\nu}}$ denotes the energy-momentum tensor of standard matter and $\mathcal{U}_{\mu\nu}$ the tensor given by (10), while:

$$\frac{1}{\sqrt{-g}} \frac{\delta(\sqrt{-g} L_{\chi_0})}{\delta g^{\mu\nu}} = \omega_0 \partial_\mu \chi_0 \partial_\nu \chi_0 - \frac{\omega_0}{2} g_{\mu\nu} \left[ \partial_\sigma \chi_0 \partial^\sigma \chi_0 + 1 \right] \tag{34}$$

$$\frac{1}{\sqrt{-g}} \frac{\delta(\sqrt{-g} L_{\chi_1})}{\delta g^{\mu\nu}} = [\omega_1 - m^2] \left[ \partial_\mu \chi_1 \partial_\nu \chi_1 - \frac{1}{2} g_{\mu\nu} \partial_\sigma \chi_1 \partial^\sigma \chi_1 \right] - \frac{\omega_1}{2} g_{\mu\nu} \tag{35}$$

$$\frac{1}{\sqrt{-g}} \frac{\delta(\sqrt{-g} L_{\chi_2})}{\delta g^{\mu\nu}} = [\omega_2 + m^2] \left[ \partial_\mu \chi_2 \partial_\nu \chi_2 - \frac{1}{2} g_{\mu\nu} \partial_\sigma \chi_2 \partial^\sigma \chi_2 \right] - \frac{\omega_2}{2} g_{\mu\nu} \tag{36}$$

$$\frac{1}{\sqrt{-g}} \frac{\delta(\sqrt{-g} \frac{\chi_2}{m} [\mathcal{E}_{\mu\nu} \mathcal{U}^{\mu\nu}]^2)}{\delta g^{\mu\nu}} = \frac{\chi_2}{m} \left[ 2 \left[ \mathcal{U}^{\mu\nu} \frac{\delta \mathcal{E}_{\mu\nu}}{\delta g^{\mu\nu}} + \mathcal{E}_{\mu\nu} \frac{\delta \mathcal{U}^{\mu\nu}}{\delta g^{\mu\nu}} \right] - \frac{1}{2} g_{\mu\nu} \mathcal{E}_{\rho\sigma} \mathcal{U}^{\rho\sigma} \right] \mathcal{E}_{\rho\sigma} \mathcal{U}^{\rho\sigma} \tag{37}$$

b) varying (28) with respect to $\delta \varphi^\rho$; $\delta \omega_0$ and $\delta \chi_0$ would respectively yield:

$$m^2 \nabla_\mu \left[ \frac{\partial \mathcal{U}(g, \varphi^\rho)}{\partial (\partial_\mu \varphi^\rho)} \right] + 2 \nabla^\mu \left[ \frac{\chi_2}{m} \mathcal{E}_{\mu\nu} \mathcal{U}^{\mu\nu} \left[ \mathcal{U}^{\mu\nu} \frac{\partial \mathcal{E}_{\mu\nu}}{\partial (\partial^\mu \varphi^\rho)} + \mathcal{E}_{\mu\nu} \frac{\partial \mathcal{U}^{\mu\nu}}{\partial (\partial^\mu \varphi^\rho)} \right] \right] = 0 \tag{38}$$

$$\partial_\mu \chi_0 \partial^\mu \chi_0 + 1 + 2 \frac{\chi_2}{m} \mathcal{E}_{\mu\nu} \mathcal{U}^{\mu\nu} \mathcal{U}^{\mu\nu} \frac{\partial \mathcal{E}_{\mu\nu}}{\partial \omega_0} = 0 \tag{39}$$

$$\nabla^\mu \left[ 2\omega_0 \partial_\mu \chi_0 + 2\frac{\chi_2}{m} \mathcal{E}_{\mu\nu} \mathcal{U}^{\mu\nu} \mathcal{U}^{\mu\nu} \frac{\partial \mathcal{E}_{\mu\nu}}{\partial(\partial^\mu \chi_0)} \right] = 0 \tag{40}$$

**c)** varying (28) with respect to $\delta\omega_1$, $\delta\varepsilon_1$ and $\delta\chi_1$ would respectively yield:

$$\partial_\mu \chi_1 \partial^\mu \chi_1 + 1 + \varepsilon_1 = 0 \tag{41}$$

$$\omega_1 - m^2 = 0 \tag{42}$$

$$\nabla^\mu \left[ 2(\omega_1 - m^2)\partial_\mu \chi_1 \right] = 0 \tag{43}$$

**d)** varying (28) with respect to $\delta\omega_2$, $\delta\varepsilon_2$, $\delta\chi_2$ and $\delta A_\mu$ would respectively yield:

$$\partial_\mu \chi_2 \partial^\mu \chi_2 - 1 + \varepsilon_2 = 0 \tag{44}$$

$$\omega_2 + m^2 = 0 \tag{45}$$

$$\nabla^\mu \left[ 2(\omega_2 + m^2)\partial_\mu \chi_2 \right] - \frac{1}{m} \left[ \mathcal{E}_{\mu\nu} \mathcal{U}^{\mu\nu} \right]^2 = 0 \tag{46}$$

$$\frac{\chi_2}{m} \mathcal{E}_{\mu\nu} \mathcal{U}^{\mu\nu} \mathcal{U}^{\mu\nu} \frac{\partial \mathcal{E}_{\mu\nu}}{\partial A_\mu} = 0 \tag{47}$$

From there, it is clear that because of (45); the above Eq. (46) would enforce:

$$\frac{1}{m} \left[ \mathcal{E}_{\mu\nu} \mathcal{U}^{\mu\nu} \right]^2 = 0 \tag{48}$$

Now since $m$ is a constant and that $\mathcal{U}^{\mu\nu}$ is not always zero; then one infers that the above (48) enforces in its turn:

$$\mathcal{E}_{\mu\nu} = \omega_0 \partial_\mu \chi_0 \partial_\nu \chi_0 - m^2 \mathcal{U}_{\mu\nu} - k(\Phi)A_\mu A_\nu - k(\Phi)S_{\mu\nu}(g, \Psi, \varphi^\rho, A) + m^2 \mathcal{V}_{\mu\nu}(\Phi) = 0 \tag{49}$$

Clearly, because of this above constraint (49); the Eqs. (38)–(40) and (47) would respectively simplify as:

$$m^2 \nabla_\mu \left[ \frac{\partial \mathcal{U}(g, \varphi^\rho)}{\partial(\partial_\mu \varphi^\rho)} \right] = 0 \tag{50}$$

$$\partial_\mu \chi_0 \partial^\mu \chi_0 + 1 = 0 \tag{51}$$

$$\nabla^\mu \left[ 2\omega_0 \partial_\mu \chi_0 \right] = 0 \tag{52}$$

$$0 = 0 \tag{53}$$

In the same way, because of (42), (45), (49) and (51); the Eqs. (34)–(37) would simplify as:

$$\frac{1}{\sqrt{-g}} \frac{\delta(\sqrt{-g}L_{\chi_0})}{\delta g^{\mu\nu}} = \omega_0 \partial_\mu \chi_0 \partial_\nu \chi_0 \tag{54}$$

$$\frac{1}{\sqrt{-g}} \frac{\delta(\sqrt{-g}L_{\chi_1})}{\delta g^{\mu\nu}} = -\frac{1}{2} m^2 g_{\mu\nu} \tag{55}$$

$$\frac{1}{\sqrt{-g}} \frac{\delta(\sqrt{-g}L_{\chi_2})}{\delta g^{\mu\nu}} = +\frac{1}{2} m^2 g_{\mu\nu} \tag{56}$$

$$\frac{1}{\sqrt{-g}} \frac{\delta(\sqrt{-g}\frac{\chi_2}{m}[\mathcal{E}_{\mu\nu}\mathcal{U}^{\mu\nu}]^2)}{\delta g^{\mu\nu}} = 0 \tag{57}$$

whence the gravitational Eq. (33) would simplify as:

$$G_{\mu\nu} = \kappa T_{\mu\nu} - m^2 \mathcal{U}_{\mu\nu} + \omega_0 \partial_\mu \chi_0 \partial_\nu \chi_0 \tag{58}$$

From there, since the constraint (49) implies also:

$$\omega_0 \partial_\mu \chi_0 \partial_\nu \chi_0 = m^2 \mathcal{U}_{\mu\nu} + k(\Phi)A_\mu A_\nu + k(\Phi)S_{\mu\nu}(g, \Psi, \varphi^\rho, A) - m^2 \mathcal{V}_{\mu\nu}(\Phi) \tag{59}$$

then we will finally get the gravitational equation reading:

$$G_{\mu\nu} = \kappa T_{\mu\nu} - m^2 \mathcal{V}_{\mu\nu}(\Phi) + k(\Phi)A_\mu A_\nu + k(\Phi)S_{\mu\nu}(g, \Psi, \varphi^\rho, A) \tag{60}$$

while the equation of motion of the Stückelberg scalar fields $\varphi^\rho$ still remains (50). Also, it is worth mentioning that because of the enforced constraint (49); we can verify that the Euler-Lagrange equation of motion of any given field $\Psi$ of the standard model of particle physics (including the Higgs field $\Phi$) would remain the usual one:

$$\nabla^\mu \left[ 2\kappa \frac{\partial L_M}{\partial(\partial^\mu \Psi)} - \frac{\chi_2}{m} \frac{\partial [\mathcal{E}_{\mu\nu} \mathcal{U}^{\mu\nu}]^2}{\partial(\partial^\mu \Psi)} \right] - 2\kappa \frac{\partial L_M}{\partial \Psi} + \frac{\chi_2}{m} \frac{\partial [\mathcal{E}_{\mu\nu} \mathcal{U}^{\mu\nu}]^2}{\partial \Psi} = \nabla^\mu \left[ 2\kappa \frac{\partial L_M}{\partial(\partial^\mu \Psi)} \right] - 2\kappa \frac{\partial L_M}{\partial \Psi} = 0 \qquad (61)$$

i.e. standard model of particle physics remains roughly unchanged in our paradigm (28).

Now, as none of the auxiliary variable $\omega_0$, $\chi_0$, $\omega_1$, $\chi_1$, $\varepsilon_1$, $\omega_2$, $\chi_2$ and $\varepsilon_2$, contribute neither in the equation of motion (61) of standard matter, nor in the equation of motion (50) for the Stückelberg fields $\varphi^\rho$, while only $A_\mu$ contribute explicitly in the gravitation Eq. (60); then excepted $A_\mu$, all the other auxiliary variables $\omega_0$, $\chi_0$, $\omega_1$, $\chi_1$, $\varepsilon_1$, $\omega_2$, $\chi_2$ and $\varepsilon_2$ could be seen as just pure mathematical fields (i.e. non-physical fields) helping to achieve the Euler-Lagrange equations of motion (50), (60) and (61).

It is also worth to mention that because of (61); applying the Bianchi identity $\nabla^\mu G_{\mu\nu} = 0$ to (60) would impose:

$$\nabla^\mu \left[ k(\Phi) A_\mu A_\nu + k(\Phi) S_{\mu\nu}(g, \Psi, \varphi^\rho, A) - m^2 \, \mathcal{V}_{\mu\nu}(\Phi) \right] = 0 \qquad (62)$$

which is the equation that enforces the auxiliary variable $A_\mu$; an equation which of course could also be obtained from (52) by considering (50), (51) and (59); whence the gravitational theory yielded by our action (28), would in summary be:

$$G_{\mu\nu} = \kappa T_{\mu\nu} - m^2 \, \mathcal{V}_{\mu\nu}(\Phi) + k(\Phi) A_\mu A_\nu + k(\Phi) S_{\mu\nu}(g, \Psi, \varphi^\rho, A) \qquad (63)$$

$$\nabla_\mu \left[ \frac{\partial \mathcal{U}(g, \varphi^\rho)}{\partial(\partial_\mu \varphi^\rho)} \right] = \nabla^\mu \mathcal{U}_{\mu\nu} = 0 \qquad (64)$$

$$\nabla^\mu \left[ k(\Phi) A_\mu A_\nu + k(\Phi) S_{\mu\nu}(g, \Psi, \varphi^\rho, A) - m^2 \, \mathcal{V}_{\mu\nu}(\Phi) \right] = 0 \qquad (65)$$

where of course $G_{\mu\nu}$ denotes the Einstein tensor; $\kappa$ the Einstein gravitational constant; $T_{\mu\nu} = -\frac{2}{\sqrt{-g}} \frac{\delta(\sqrt{-g} L_M)}{\delta g^{\mu\nu}}$ the energy-momentum tensor of standard matter; $m$ a free constant; $k(\Phi)$ the function given by (20); $A_\mu$ an auxiliary variable; $\mathcal{U}_{\mu\nu}$ the tensor given by (10); $\mathcal{V}_{\mu\nu}(\Phi)$ the tensor given by (27), $S_{\mu\nu}(g, \Psi, \varphi^\rho, A)$ a free symmetric tensor that can depend on the standard matter fields, the metric, the Stückelberg fields $\varphi^\rho$ and/or their derivatives etc…, as well as of the auxiliary vector field $A_\mu$.

Now, what is interesting with (63)–(65), is that, it is not only a "gravitational theory" derivable from variational principle but also, it is (as we are going to see in the next section), the first example of four-dimensional modified gravity which is massless before electroweak phase transition and massive after electroweak phase transition i.e. a gravity theory for which the graviton gain mass dynamically through electroweak spontaneous symmetry breaking; let's see how.

## 4. Massive after electroweak phase transition, massless before

### 4.1. After electroweak phase transition

To verify that our gravitational theory (63)–(65) is a massive gravity theory after the electroweak phase transition; it suffices to notice that the Eq. (20) garantees that after the electroweak phase transition we have:

$$k(\Phi) = 0 \qquad (66)$$

As consequence, after the electroweak phase transition we will have (see *subsection 2.3.*):

$$\mathcal{V}_{\mu\nu}(\Phi) = \mathcal{U}_{\mu\nu} \qquad (67)$$

whence after the electroweak phase transition where $k(\Phi) = 0$, the equations (63–65) will transform/simplify into:

$$G_{\mu\nu} = \kappa T_{\mu\nu} - m^2 \mathcal{U}_{\mu\nu} \qquad (68)$$

$$\nabla_\mu \left[ \frac{\partial \mathcal{U}(g, \varphi^\rho)}{\partial(\partial_\mu \varphi^\rho)} \right] = \nabla^\mu \mathcal{U}_{\mu\nu} = 0 \qquad (69)$$

This latter (68-69) being nothing than exactly the dRGT massive gravity (8-9); then we can quite affirm that after the electroweak phase transition, our gravitational theory (63–65) yields/becomes explicitly a massive gravity, more precisely the famous dRGT massive gravity [5,10].

Let's note that although (63–65) becomes dRGT massive gravity through the eletroweak phase transition; the mass $m$ of the graviton in (68-69) is a free constant completely independent from the electroweak scale or parameter i.e. $m$ is not constrained in any way by electroweak parameter, it is really a free parameter in the theory which could for example be very small and thus compatible with experimental limit [14]; likewise, the evolution of $k(\Phi)$ which not appears in (68-69) is not affected by massive gravity instead it principally depends from the field $\Phi$ and its vacuum expectation value.

## 4.2. Before electroweak phase transition

Now if as seen in the previous subsection, (63-65) yields/becomes after the electroweak phase transition, a massive gravity, more precisely the famous dRGT massive gravity (8-9); we are going to see in this subsection that before the electroweak phase transition, (63-65) is "well and truly" a massless gravity theory.

Indeed as before the electroweak phase transition; we have $\Phi_1 \neq \Phi_2$; then it results that before electroweak phase transition, we have:

$$k(\Phi) \neq 0 \tag{70}$$

and thus:

$$\mathcal{U}_{\mu\nu}(\Phi) \neq \mathcal{U}_{\mu\nu} \tag{71}$$

Better, $\mathcal{U}_{\mu\nu}(\Phi)$ will in this case $k(\Phi) \neq 0$, be (see *subsection 2.3.*) a tensor incapable of providing a fixed reference background metric alongside the physical spacetime metric $g_{\mu\nu}$, and thus a tensor incapable of yielding mass-term for the graviton.

This can be verified more easily by linearizing (63-65). Indeed as well known [8]; in the linearized regime characterized by $g_{\mu\nu} = \eta_{\mu\nu} + h_{\mu\nu}$, with $h_{\mu\nu} \ll 1$ a perturbation around the Minkowski metric $\eta_{\mu\nu}$; the Einstein tensor $G_{\mu\nu}$ could be simplified as [8]:

$$G_{\mu\nu} = -\frac{1}{2}\Box \tilde{h}_{\mu\nu} \tag{72}$$

where $\Box = \nabla_\mu \nabla^\mu$ denotes the d'Alembert operator and $\tilde{h}_{\mu\nu} = h_{\mu\nu} - \frac{1}{2}h\eta_{\mu\nu}$ the trace-reversed perturbation variable (*graviton*).

whence in the linearized regime ($g_{\mu\nu} = \eta_{\mu\nu} + h_{\mu\nu}$ with $h_{\mu\nu} \ll 1$); our gravity Eq. (63) would simplify as:

$$\Box \tilde{h}_{\mu\nu} = -2\kappa T_{\mu\nu} + 2m^2 \, \mathcal{U}_{\mu\nu}(\Phi) - 2k(\Phi) A_\mu A_\nu - 2k(\Phi) S_{\mu\nu}(g, \Psi, \varphi^\rho, A) \tag{73}$$

From there let's for simplicity, consider case $\alpha_3 = \alpha_4 = 0$ and $f_{\rho\beta} = \eta_{\rho\beta}$ i.e. the fudicial metric is the Minkowki metric.

In this case, it is well known that the tensor $\mathcal{U}_{\mu\nu}$ would in the linearized regime ($g_{\mu\nu} = \eta_{\mu\nu} + h_{\mu\nu}$ with $h_{\mu\nu} \ll 1$), simplify as [5,9,10]:

$$\mathcal{U}_{\mu\nu} = \frac{1}{2} \tilde{h}_{\mu\nu} \tag{74}$$

Now since we have $\mathcal{U}_{\mu\nu}(\Phi) = \mathcal{U}_{\mu\nu}$ when $k(\Phi) = 0$; then it results that in regime $k(\Phi) = 0$ i.e. after the electroweak phase transition; the Eq. (73) for the graviton would become:

$$\left[\Box - m^2\right] \tilde{h}_{\mu\nu} = -2\kappa T_{\mu\nu} \tag{75}$$

which is nothing than the equation of motion of a massive graviton of mass $m$; whence as already discussed in the previous subsection, our model (63-65) is after the electroweak phase transition where $k(\Phi) = 0$, a massive gravity theory, more precisely the famous dRGT massive gravity [5,9,10]. However since in the regime $k(\Phi) \neq 0$ i.e. before the electroweak phase transition, the tensor $\mathcal{U}_{\mu\nu}(\Phi)$ could for $\alpha_3 = \alpha_4 = 0$, write:

$$\mathcal{U}_{\mu\nu}(\Phi) = g_{\mu\nu}\left[6 + 2f_\varphi \, \Phi - f_\varphi \sqrt{\Phi_\rho^\sigma \Phi_\sigma^\rho}\right] - f_\varphi \, \Phi_{\mu\nu} - \left[3 - 2\sqrt{f_\varphi \, \Phi}\right]\sqrt{f_\varphi g_{\mu\nu} \, \Phi_{\mu\nu}} \tag{76}$$

where $f_\varphi = \partial_\rho \varphi^\rho \partial_\rho \varphi^\rho$, with $\varphi^\rho$ the four Stückelberg scalars fields; $\Phi = g^{\mu\nu} \Phi_{\mu\nu}$ with $\Phi_{\mu\nu}$ the tensor given by (25).

then it results that before the electroweak phase transition where $k(\Phi) \neq 0$; the Eq. (73) for the graviton would read:

$$\Box \tilde{h}_{\mu\nu} = -2\kappa T_{\mu\nu} - 2m^2 \, \mathcal{T}_{\mu\nu} \tag{77}$$

where:

$$\mathcal{T}_{\mu\nu} = \frac{k(\Phi)}{m^2}\left[A_\mu A_\nu + S_{\mu\nu}(g, \Psi, \varphi^\rho, A)\right] + f_\varphi \, \Phi_{\mu\nu} - \left[6 + 2f_\varphi \, \Phi - f_\varphi \sqrt{\Phi_\rho^\sigma \Phi_\sigma^\rho}\right]g_{\mu\nu} + \left[3 - 2\sqrt{f_\varphi \, \Phi}\right]\sqrt{f_\varphi g_{\mu\nu} \, \Phi_{\mu\nu}} \tag{78}$$

Now since neither $\mathcal{A}_{\mu\nu}$ nor $\Phi_{\mu\nu}$ are fudicial metrics or involve fudicial metrics when $k(\Phi) \neq 0$; then it results that the Eq. (77) describes "well and truly" a massless graviton $\tilde{h}_{\mu\nu}$, since neither the energy-momentum tensor $T_{\mu\nu}$ of standard matter, nor the extra *energy-momentum tensor* $\mathcal{T}_{\mu\nu}$ carry/yield mass-term for the graviton $\tilde{h}_{\mu\nu}$; whence our model (63-65) is in the regime $k(\Phi) \neq 0$ i.e. before the electroweak phase transition, a massless gravity theory, more precisely, a modified massless gravity theory departing from General Relativity [13] by the presence of the tensor $\mathcal{T}_{\mu\nu}$ given by (78).

It is worth mentioning that the four Stückelberg scalar fields $\varphi^\rho$ because of their equation of motion (64) carry only three propagating physical degrees of freedom [5,9,10]; meaning that the tensor $\mathcal{T}_{\mu\nu}$ carries only three additional propagating physical degrees of freedom since the auxiliary variable $A_\mu$ enters non-dynamically (i.e. without kinetic term) in the action (28) and thus not gives rise to additional propagating physical degrees of freedom.

This mean that, both before the electroweak phase transition where it is a massless gravity, as well as after the electroweak phase transition where it is a massive gravity, our model (63-65) has five propagating physical degrees of freedom if we not count the degrees of freedom carried by $T_{\mu\nu}$; the two yielded by the metric tensor $g_{\mu\nu}$ plus the three yielded by the four Stückelberg scalar fields $\varphi^\rho$ [5,9,10]; whence we can say that the graviton of our gravitational theory (63-65) gains mass (dynamically and in a consistent way) through the dynamical standard model spontaneous electroweak symmetry breaking, and thus that we achieved a consistent dynamical Higgs mechanism for dRGT gravity.

## 5. Strong coupling break-down problem resolution

The principal interest of the dynamical higgs mechanism for gravity that we propose here is its ability to circumvent the strong coupling break-down problem of massive gravity.
Indeed, the fact that (63-65) describes a massless gravity theory which becomes dynamically massive after electroweak phase transition means that the gotten dRGT massive gravity would escape to strong coupling break down problem if the mass $m$ of its graviton is at least of the electroweak scale $\Lambda_{EW}$ (temperature of electroweak symmetry breaking) i.e. if $m \geq \Lambda_{EW} \sim 160 GeV$, insofar as in this case, the cutoff-scale $\Lambda_3 = [m^2 M_P]^{\frac{1}{3}}$ beyond which the gotten dRGT massive gravity breaks-down because of strong coupling, will be at/or beyond the electroweak scale $\Lambda_{EW}$, a scale beyond which the considered dRGT massive gravity becomes the massless gravity reading:

$$G_{\mu\nu} = \kappa T_{\mu\nu} + m^2 \, \mathcal{T}_{\mu\nu} \tag{79}$$

$$\nabla_\mu \left[ \frac{\partial \mathcal{U}(g,\varphi^\rho)}{\partial(\partial_\mu \varphi^\rho)} \right] = \nabla^\mu \mathcal{U}_{\mu\nu} = 0 \tag{80}$$

$$\nabla^\mu \mathcal{T}_{\mu\nu} = 0 \tag{81}$$

where of course $\mathcal{T}_{\mu\nu}$ denotes the tensor given by (78).

In other words, in the case of $m \geq \Lambda_{EW} \sim 160 GeV$; the dRGT massive gravity (68-69) yielded by (63-65) would become the above massless gravity (79–81) before going beyond the scale $\Lambda_3 = [m^2 M_P]^{\frac{1}{3}}$ where it breaks-down because of strong coupling. Said otherwise, (63-65) yield a dRGT massive gravity which would escape to the "strong coupling break down problem of massive gravity" if the graviton has a mass $m \geq \Lambda_{EW}$, since in this case ($m \geq \Lambda_{EW}$), the gotten dRGT massive gravity whould before entering in its strong coupling regime where it breaks-down, become the massless gravity (79–81) which, as a massless gravity could, depending from the exact expression of $S_{\mu\nu}(g, \Psi, \varphi^\rho, A)$, have a cutoff-scale of the order of Planck Mass $M_P \sim 10^{19} GeV$ if not being UV complete.

Unfortunately, inspite this step forward, things are not so satisfying, insofar as (63-65) is able to overcome the strong coupling break-downn problem of dRGT massive gravity only if we assume for the graviton a mass $m \geq \Lambda_{EW} \sim 160 GeV$ whilts experimental data exclude for the graviton a mass $m > 6.76 \times 10^{-23} eV$ [14].

In other words, we cannot with (63-65) have in the same time a dRGT massive gravity compatible with experimental data and avoiding strong coupling break-down problem since so that to avoid the strong coupling break-down problem, the graviton should have a mass $m \geq \Lambda_{EW} \sim 160 GeV$ while so that to be compatible with experimental data, the graviton should have a mass $m < 6.76 \times 10^{-23} eV$ [14].

However it is important to understand that this inability of the dRGT gravity yielded by (63-65) of being in the same time compatible with experimental data and avoid strong coupling break-down problem comes only from the fact that the electroweak scale $\Lambda_{EW} \sim 160 GeV$ at which the transition between the massless gravity (79–81) and the massive gravity (68-69) takes place through electroweak symmetry breaking/restoration is too high and far beyond the strong coupling break-down scale of a dRGT massive gravity of experimentally viable graviton mass $m < 6.76 \times 10^{-23} eV$ [14]. In reality to have a dRGT massive gravity capable of avoiding strong coupling break-down problem and having a graviton of for example mass $m \sim 10^{-33} eV$ compatible with experimental data; we could/should at least:

-on the one hand, add in the action (28), for example the following lagrangian $L(\phi_1, \phi_2)$ of a real scalar field $\phi_1$ coupled to a complex SU(2) doublet scalar field $\phi_2$, reading:

$$L(\phi_1,\phi_2) = -\partial_\mu \phi_1 \partial^\mu \phi_1 - \partial_\mu \phi_2^\dagger \partial^\mu \phi_2 - \lambda_1 [\phi_1]^2 \phi_2 \phi_2^\dagger - m_2 \phi_2^\dagger \phi_2 - \lambda_2 [\phi_2^\dagger \phi_2]^2 \tag{82}$$

where $\lambda_1 > 0$, $\lambda_2 > 0$ and $m_2 < 0$ are free constants.

-while on the other hand, we should replace into the action (28); the standard model Higgs field function $k(\Phi)$ given by (20) by the following reading:

$$k(\phi_2) = \left[ 2B(e^{\phi_1}, 1 + e^{\phi_2}) - B(e^{\phi_2}, e^{\phi_2}) \right]^2 \tag{83}$$

where e is the Euler's number or Napier's constant; $B(x,y) = \frac{\Gamma(x)\Gamma(y)}{\Gamma(x+y)}$ the beta function; while $\phi_1 = \frac{\sqrt{v^2 - [2v - \sqrt{2\phi_2^\dagger \phi_2}]\sqrt{\phi_2^{\dagger*}\phi_2 + \phi_2^\dagger \phi_2^*}}}{4v}$ and $\phi_2 = \frac{\sqrt{2\phi_2^\dagger \phi_2}}{4v} - \frac{1}{4}$, with $v = \sqrt{-\frac{m_2}{\lambda_2}}$ a constant; $\phi_2^*$ the complex conjugate of $\phi_2$ and $\phi_2^{\dagger*}$ the complex conjugate of $\phi_2^\dagger$.

Of course, we can verify that by constuction of $\phi_1$ and $\phi_2$; it is only when the SU(2) symmetry associated to the field $\phi_2$ is sponaneously broken (like in the case where $\phi_2 = \frac{1}{\sqrt{2}}[v + h]$) that we could have:

$$\phi_1 = \phi_2 = \frac{h}{4v} \tag{84}$$

whence in practice, we will have $k(\phi_2) = 0$ only in the case where $\phi_2$ breaks the full SU(2) symmetry of $L(\phi_1,\phi_2)$ by for example developing the non-zero vacuum expectation value $\frac{v}{\sqrt{2}} = \sqrt{-\frac{m_2}{2\lambda_2}}$; a symmetry breaking which can takes place at very low energy scale depending on the values of the free constants $\lambda_1 > 0$, $\lambda_2 > 0$ and $m_2 < 0$. Now as one can verify; by considering $L(\phi_1, \phi_2)$ and $k(\phi_2)$ into (28), and replacing $k(\Phi)$ by $k(\phi_2)$, we would insead of the theory (63-65), get the one reading:

$$G_{\mu\nu} = \kappa T_{\mu\nu} - m^2 \, \mathcal{U}_{\mu\nu}(\phi_2) + k(\phi_2)\left[A_\mu A_\nu + S_{\mu\nu}(g, \Psi, \varphi^\rho, A)\right] + T^{\phi_1}_{\mu\nu} + T^{\phi_2}_{\mu\nu} + \frac{V(\phi)}{2} g_{\mu\nu} \tag{85}$$

$$\nabla_\mu \left[\frac{\partial \mathcal{U}(g, \varphi^\rho)}{\partial(\partial_\mu \varphi^\rho)}\right] = \nabla^\mu \mathcal{U}_{\mu\nu} = 0 \tag{86}$$

$$\nabla^\mu \left[k(\phi_2) A_\mu A_\nu + k(\phi_2) S_{\mu\nu}(g, \Psi, \varphi^\rho, A) - m^2 \, \mathcal{U}_{\mu\nu}(\phi_2)\right] = 0 \tag{87}$$

$$\Box \phi_1 - \lambda_1 \phi_1 \phi_2 \phi_2^\dagger = 0 \tag{88}$$

$$\Box \phi_2 - \lambda_1 [\phi_1]^2 \phi_2 - m_2 \phi_2 - 2\lambda_2 [\phi_2 \phi_2^\dagger] \phi_2 = 0 \tag{89}$$

$$\Box \phi_2^\dagger - \lambda_1 [\phi_1]^2 \phi_2^\dagger - m_2 \phi_2^\dagger - 2\lambda_2 [\phi_2 \phi_2^\dagger] \phi_2^\dagger = 0 \tag{90}$$

where of course $G_{\mu\nu}$ denotes the Einstein tensor; $\kappa$ the Einstein gravitational constant; $T_{\mu\nu}$ the energy-momentum tensor of standard matter; $m$, $\lambda_1$, $\lambda_2$, $m_2$ free constants; $k(\phi_2)$ the function given by (83); $A_\mu$ an auxiliary variable; $\mathcal{U}_{\mu\nu}$ the tensor given by (10); $\mathcal{U}_{\mu\nu}(\phi_2)$ the tensor obtained by replacing into $\mathcal{U}_{\mu\nu}(\Phi)$ the function $k(\Phi)$ by the $k(\phi_2)$ given by (83); $S_{\mu\nu}(g, \Psi, \varphi^\rho, A)$ a free symmetric tensor that can depend on the standard matter fields, the metric, the Stückelberg field $\varphi^\rho$ and/or their derivatives etc…, as well as of the auxiliary vector field $A_\mu$, while $T^{\phi_1}_{\mu\nu} = \partial_\mu \phi_1 \partial_\nu \phi_1 - \frac{1}{2} g_{\mu\nu} \partial_\sigma \phi_1 \partial^\sigma \phi_1$; $T^{\phi_2}_{\mu\nu} = \partial_\mu \phi_2^\dagger \partial_\nu \phi_2 - \frac{1}{2} g_{\mu\nu} \partial_\sigma \phi_2^\dagger \partial^\sigma \phi_2$ and $V(\phi) = \lambda_1 [\phi_1]^2 \phi_2^\dagger \phi_2 + m_2 \phi_2^\dagger \phi_2 + \lambda_2 [\phi_2^\dagger \phi_2]^2$.

From there, it is easy to verify that when $\phi_2$ breaks the SU(2) symmetry i.e. when for example $\phi_2$ develops the non-zero vacuum expectation value $\frac{v}{\sqrt{2}} = \sqrt{-\frac{m_2}{2\lambda_2}}$, we would instead of the standard dRGT massive gravity (68-69), get the one reading:

$$G_{\mu\nu} = \kappa T_{\mu\nu} - m^2 \mathcal{U}_{\mu\nu} + T^{\phi_1}_{\mu\nu} + T^{\phi_2}_{\mu\nu} + \frac{V(\phi)}{2} g_{\mu\nu} \tag{91}$$

$$\nabla_\mu \left[\frac{\partial \mathcal{U}(g, \varphi^\rho)}{\partial(\partial_\mu \varphi^\rho)}\right] = \nabla^\mu \mathcal{U}_{\mu\nu} = 0 \tag{92}$$

$$\Box \phi_1 - \lambda_1 \phi_1 \phi_2 \phi_2^\dagger = 0 \tag{93}$$

$$\Box \phi_2 - \lambda_1 [\phi_1]^2 \phi_2 - m_2 \phi_2 - 2\lambda_2 [\phi_2 \phi_2^\dagger] \phi_2 = 0 \tag{94}$$

with (93) and (94) *describing in the details two massive and three massless (Goldstone) scalar fields*.

while before the spontaneous breaking of the SU(2) symmetry associated to $\phi_2$, we would get the massless gravity theory reading:

$$G_{\mu\nu} = \kappa T_{\mu\nu} + m^2 \, \mathcal{T}_{\mu\nu}(\phi_2) + T^{\phi_1}_{\mu\nu} + T^{\phi_2}_{\mu\nu} + \frac{V(\phi)}{2} g_{\mu\nu} \tag{95}$$

$$\nabla_\mu \left[\frac{\partial \mathcal{U}(g, \varphi^\rho)}{\partial(\partial_\mu \varphi^\rho)}\right] = \nabla^\mu \mathcal{U}_{\mu\nu} = 0 \tag{96}$$

$$\nabla^\mu \, \mathcal{T}_{\mu\nu}(\phi_2) = 0 \tag{97}$$

$$\Box \phi_1 - \lambda_1 \phi_1 \phi_2 \phi_2^\dagger = 0 \tag{98}$$

$$\Box \phi_2 - \lambda_1 [\phi_1]^2 \phi_2 - m_2 \phi_2 - 2\lambda_2 [\phi_2 \phi_2^\dagger] \phi_2 = 0 \tag{99}$$

$$\Box \phi_2^\dagger - \lambda_1 [\phi_1]^2 \phi_2^\dagger - m_2 \phi_2^\dagger - 2\lambda_2 [\phi_2 \phi_2^\dagger] \phi_2^\dagger = 0 \tag{100}$$

where of course $\mathcal{T}_{\mu\nu}(\phi_2)$ denotes the tensor obtained by replacing into $\mathcal{T}_{\mu\nu}$ the function $k(\Phi)$ by the $k(\phi_2)$ given by (83).

Now, what is interesting with the model (85–90), is that, in contrary to (63–65); the dRGT massive gravity (91–94) that it yields could be compatible with experimental data (i.e. having a graviton of mass $m < 6.76 \times 10^{-23} eV$) while escaping to strong coupling break down problem.

This is due to the fact that in the case of (85–90), the dynamical symmetry breaking what leads to (91–94) could take place at the low energy-scale $\Lambda_\phi$ that we want as long as we assume appropriate values for the constants $\lambda_1$, $\lambda_2$, $m_2$.

As consequence, no matter the exact value of the constant $m > 0$; we could choose $\lambda_1$, $\lambda_2$, $m_2$ in a way guaranteeing that the cutoff-scale $\Lambda_3 = [m^2 M_P]^{\frac{1}{3}}$ above which (91–94) breaks-down because of strong coupling, be beyond the scale $\Lambda_\phi$ where (91–94) gives way to the massless gravity (95–100).

In other words, we could, by choosing appropriate values for the constants $\lambda_1$, $\lambda_2$, $m_2$, always place the scale $\Lambda_3 = [m^2 M_P]^{\frac{1}{3}}$ (beyond which (91–94) breaks-down because of strong coupling) above the scale $\Lambda_\phi$ beyond which (91–94) gives way to (95–100); said otherwise, no matter the exact value of $m > 0$, we could always make (91–94) escaping to massive gravity strong coupling breakdown problem by choosing for $\lambda_1$, $\lambda_2$, $m_2$ values guaranteeing that (91–94) gives way to the massless gravity (95–100) before entering in the strong coupling regime where it breaks-down.

Now since as a massless gravity theory, the model (95–100) could, depending from the exact expression of $S_{\mu\nu}(g, \Psi, \varphi^\rho, A)$, have a cutoff-scale of the order of Planck Mass $M_P \sim 10^{19} GeV$ if not being UV complete; then one concludes that with (85–90) we could, through dynamical spontaneous symmetry breaking, get a dRGT massive gravity capable of escaping to strong coupling coupling break-down problem while having a graviton mass $m < 6.76 \times 10^{-23} eV$ compatible with experimental data [14]; whence with (63–65) and more exactly with (85–90), we succeeded in achieving the first example of dynamical Higgs mehanism for gravity permitting to overcome the strong coupling break-down problem of massive gravity.

Before concluding, it is worth mentioning that the trick applied in the present work to "dRGT massive gravity" in order to get a dynamical Higgs mechanism for gravity permitting to circumvent the strong coupling break-down problem of massive gravity, could be also applied or even be more naturally applied to mass-varying massive gravity [15] as we will see in a next paper.

Let's also note that it would be interesting to study more in details the massless gravity (95–100) in order to compare it with general relativity [8]; this would be the aim of a futur paper.

## 6. Conclusion

In this paper, we succeeded in showing that, it is clearly possible to achieve from variational principle a four-dimensional massless gravity theory (63–65) which becomes dynamically massive through electroweak phase transition.

After having seen that the dRGT massive gravity given by (63–65), is capable of overcoming strong coupling break-down problem only for graviton of mass $m \geq \Lambda_{EW} \sim 160 GeV$; we proposed a better model (85–90) yielding (trough dynamical symmetry breaking), the dRGT massive gravity (91–94) capable of overcoming strong coupling break-down problem for any graviton mass $m > 0$ as long as we assume appropriate values for the constants $\lambda_1, \lambda_2, m_2$.

In other words, we succeeded in achieving among the first example of dynamical Higgs mechanism for dRGT gravity permitting to overcome the strong coupling break-down problem of massive gravity and thus that we succeeded in getting among the first example of modified gravity (85–90) capable of yielding a massive gravity free from both ghost and strong coupling break-down problems.

**CRediT authorship contribution statement**

**Emmanuel Kanambaye:** Conceptualization, Data curation, Formal analysis, Funding acquisition, Investigation, Methodology, Project administration, Resources, Supervision, Validation, Writing – original draft, Writing – review & editing.

**Data availability**

No data was used for the research described in the article.

**Declaration of competing interest**

The authors declare that they have no known competing financial interests or personal relationships that could have appeared to influence the work reported in this paper.